\documentstyle[11pt,aaspp4]{article}
\def\const{\mbox{const}}
\begin{document}

\title{ GRBs: when do blackbody spectra look like non-thermal ones?}
\author{S.I.Blinnikov, A.V.Kozyreva, and
I.E.Panchenko\footnotemark\\
Institute for Theoretical and Experimental Physics,
117259 Moscow, Russia\\
Sternberg Astronomical Institute, 119899 Moscow, Russia
}

\begin{abstract}
We argue that a nonthermally looking spectrum of a gamma-ray burst
(GRB) can be formed
as a superposition of a set of thermal blackbody spectra.
This superposition may be done by time integration which is present
even in `time resolved' GRB spectroscopy.
A nonthermal spectrum can be obtained also by the space integration which
should take
place unless all the emission comes from a plane front moving exactly
towards the observer. 
We propose a model of the gamma-ray burst spectrum formation 
based on this idea.
This model allows the GRB radiation to be optically thick and to have
higher values of baryon load. Thus the latter is limited by the
energy considerations only, 
and not by the condition of a small optical depth.
\end{abstract}


\footnotetext{E-mail: blinn@sai.msu.su, sasha@sai.msu.su, ivan@sai.msu.su}

\section{Motivation}

Gamma-ray bursts (GRBs) still remain an unresolved mystery of modern
astrophysics
in spite of recent progress in the observations of their X-ray, optical and 
radio counterparts. Not only the nature of internal engine, but even the
mechanism of gamma-ray emission is unclear. 
Studying the spectra of GRBs is one of the keys that can unlock
this great mystery in future.

Observations of the GRB spectra  (Band et al. 1993)
show that, in general, they are well described by a 
low-energy power law with the exponent $\alpha$, being exponentially cut of at 
$E\sim E_0$, and by a high-energy power law with the exponent $\beta$. Though 
the  values of $(\alpha,\beta,E_0)$ can be different for individual bursts,
they usually are in the range of 
$(\sim -1.5 \dots 0.5,-3\dots 2,100\dots200\mbox{keV})$. 

Note that in this paper we consider the photon spectrum $N(E)$ or $N(\nu)$,
the differential energy flux density $F_\nu=h\nu N(h\nu)$, and $\nu F_{\nu}$
distribution. By default, all the power indices in this paper refer to $N(E)$.

The power-law appearance of the spectra can possibly be explained by
the  hypothesis of their nonthermal origin. The synchrotron shock mechanism
(Tavani 1996), where the GRB emission is produced
by  an optically thin relativistic plasma in a weak magnetic field,
is one of those models which give a good agreement with observed spectra.
Cohen et al (1997) find that the low-energy spectral index $\alpha$ 
in the {\it time-integrated}  of GRB is usually 
in the range from $-2/3$ to $-3/2$ as predicted by the synchrotron shock model.
The limits of this range correspond to 
the synchrotron spectra of instantaneous sample of electrons and 
the one integrated over their radiative decay 
(Rybicki \& Lightman 1979).

However, Crider, Liang \& Preece (1997) have shown, on the basis of the
analysis of the {\it time-resolved} spectra of 99 GRBs, that
neither the synchrotron shock nor the simple inverse Compton
mechanism can explain the instantaneous GRB spectra and their evolution:
the time-resolved spectral slope $\alpha$ is often  outside
the limits of the synchrotron model and does not change 
monotonically with time, as the inverse Compton model predicts.
 
While these models of gamma-ray bursts (which generally fit the observations)
have some difficulties to 
match them in detail, we can present here a blackbody
model that should be at least not worse than the other current ones.

The conflict of the optically thick model for GRBs with observations was
discussed already by Paczy\'nski (1986) and Goodman (1986).
Paczy\'nski (1986) mentioned: `The observed spectra are averaged over large
fractions of a second, and this may be responsible for the shallow slope of
the low energy part of the spectrum'.
The problem was raised recently by
Band \& Ford (1997). 
They have posed a question `whether burst
spectra are narrowband on short time-scales'. So, the question is: are the
observed broadband GRB spectra formed by time integrations of an 
evolving quasi-blackbody instantaneous spectrum, or not. 
Band \& Ford (1997) found no evidence for narrowband emission down
to 1 ms time-scale. In the present paper we consider  time-scales that
are shorter for an observer.

It is well known, that assuming high values of Lorentz factor $\Gamma$
of the GRB ejecta is necessary to solve the {\it compactness problem}
(Guilbert, Fabian \& Rees 1983,  Paczy\'nski 1986, Goodman 1986, 
Krolik \& Pier 1991,  Rees \&  M\'esz\'aros 1992, Piran 1996).
The typical time-scale of the variability
of the gamma-ray emission $\Delta t\sim10^{-2}$~seconds implies the size
of the emitting region $R<c\Delta t$, as small as $\sim 10^3$~km. The
enormous number of gamma photons in such a small volume should produce
electron-positron pairs which make the emitting region optically thick. 
This conflicts with the
observed nonthermal spectra unless one supposes that the emitting region moves
towards  the observer at a relativistic speed with Lorentz factor $\Gamma$,
then its size would be $\Gamma^2c\Delta t$, and the optical depth
correspondingly smaller.

We propose an important supplement to this solution of the compactness
problem. In our version, the relativistic
motion is still required, in order to provide the formation of an integrated
spectra from an ensemble of the thermal ones.

It is known that a sum of different thermal blackbody spectra
can produce a power-like spectrum looking as a nonthermal one.
It happens, e.g.,  in the classical
case of the Shakura-Sunyaev thin accretion disk 
(Shakura \& Sunyaev 1973).
As shown in this paper, a similar approach can provide an analogous result 
in the case of a relativistically moving emitter.

Evidently, in any realistic situation
the spectrum produced by an optically thick body is never a pure blackbody,
because of opacity (and hence emission) dependence on wavelength, the
effects of sphericity (see 
Mihalas 1978) etc. For us the black body is
just a `toy' model which is however far enough from the spectra of an 
optically thin plasma, invoked by others for explaining GRBs.
Ryde \& Svensson (1999) consider another basic model (a non-thermal one)
and show that the observed spectra result from the time integration. Our
approach is more radical than that.

By the spectrum formation model presented in this paper 
we do not introduce a new physical model of gamma-ray bursts.
We simply point out the fact that the observed non-thermal spectrum can be 
produced by an optically thick body. The assumptions needed for this
seem not to
be very unnatural. If such a picture can be worked out as a physical one
(not only the `toy' model), then new classes of GRB models become possible, 
producing `dirty' fireballs, e.g. by the neutrino annihilation
(Goodman, Dar, \& Nussinov 1987). On the GRB models 
 with a moderately high baryon load see Woosley (1993), Ruffert et al.
(1997), Fuller \& Shi (1998), Fryer \& Woosley (1998),
Popham, Woosley, \& Fryer (1998).

\section{The model of spectrum formation}

Let us assume that the emitting surface is moving towards the observer 
with $\Gamma\sim 10^3$  -- it can be an expanding shell, or 
a blob, or a `bullet', or an  `internal shock' (e.g. Piran 1998) 
-- and producing at each instant a pure blackbody spectrum
(which has a resemblance to the real spectra of optically thick plasmas).

\begin{figure}
\begin{center}   
\epsfysize=8cm
\epsfbox{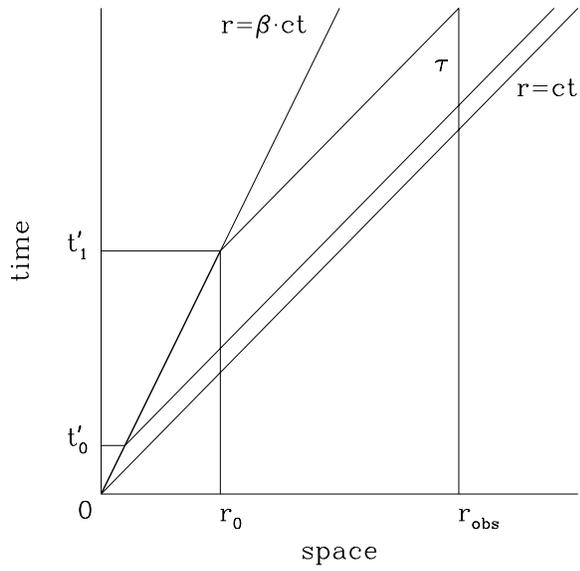}
\end{center}
\caption{
The space-time diagram for the 
emission of a shell expanding with
velocity $\beta c$.
The observer detects the duration $\tau$ for 
the pulse emitted by the shell during the interval $t'_0\ldots t'_1$.
}

\label{lorents}
\end{figure}  

Due to the well known effect, if the emitter is moving towards the observer
with the velocity $v$ corresponding to  $\Gamma=(1-v^2/c^2)^{-1/2}$ 
then the emitter and observer time-scales differ  by a factor of $2\Gamma^2$
(e.g.  Rees \&  M\'esz\'aros 1992, Shaviv \& Dar 1995,
Piran 1998, Dar 1998). 
Here and below we assume that all clocks
are synchronized in the observer's rest frame, i.e. the effect is purely
kinematical  (see Fig.\ref{lorents}), moreover it is Galilean, not 
truly relativistic (in the sense that Relativity plays no role in this effect).
The Lorentz factor $\Gamma$ is here simply a measure of the deviation
of $v$ from $c$, and nothing else.
The difference of the emitter and observer time-scales means
that, for example, $\tau=10$~ms, the time of integration by an observer,
corresponds to $\tau'\sim 5$~hours of 
emission time (Fig.\ref{lorents}). During this long time the emitting object 
can expand and cool significantly, so the spectra it produces in the beginning
and at the end of the observation interval $\tau$ can differ drastically. 
Therefore, the observed spectrum is formed by an integration of some
cooling sample of instantaneous spectra. 

For simplicity, we assume that the temperature $T$ and the area $A$ 
of the emitting object change with time as described by the following 
power laws:
\begin{equation}
T = T_0 (t/t_0)^{-\theta} = T_0 (t'/t'_0)^{-\theta} ; \qquad 
A = A_0 (t/t_0)^\sigma = A_0 (t'/t'_0)^\sigma. 
\end{equation} 
Here and below the primed time variables will refer to the emission time,
while the non-primed ones - to the detection time.

\subsection{Analytic treatment}

\begin{figure}
\begin{center}
\epsfysize=8cm
\epsfbox{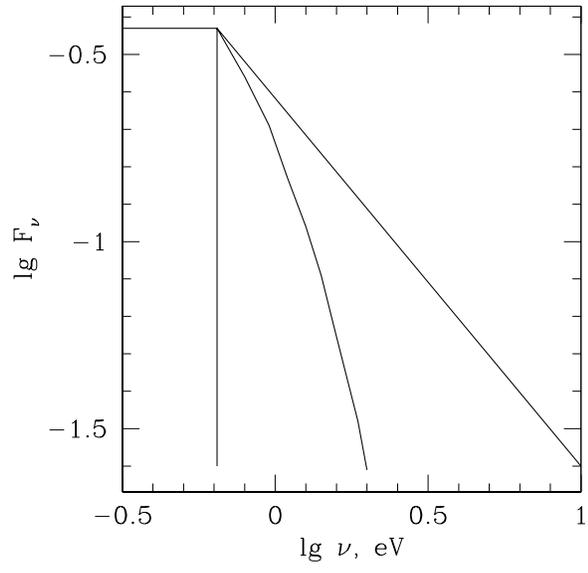}
\end{center}
\caption{The three cases of simple elementary spectra: 
 with the power (a), stepwise (b), and Plank (c) high-energy tails.}
\label{Puchok}
\end{figure}

Let us consider a set of arbitrary elementary spectra. If the members of
the set have a parameter distributed according to a power law, then the
integration of elementary spectra leads quite often to the formation
of a power spectrum. Let us show this with a simple example
(Fig.~\ref{Puchok}).

We would like to denote the elementary (instant) spectrum as $n(E)$, and the 
resulting (integral) one as $N(E)$. The spectra will be integrated in time
from $t_0$ to $t_1 = t_0 + \tau$, where $t_1 \gg t_0$.

\begin{enumerate}
\item{} Let the elementary spectrum be $n(E)\propto E^{-\beta}$ 
(Fig.~\ref{Puchok}a) in the high energy
part of the spectrum ($E>E_0$) and constant if ($E<E_0$),
where $E_0$ evolves in time like $t^{-\theta}$ and at $t_1$ reaches
the value $E_1 = E_0 (t_1/t_0)^{-\theta} $.
Then the observed integral spectrum should be
\begin{equation}
 N(E) = \int\limits_{t_0}^{t_1} A(t) n(E,t) dt  \sim
   \left\{\begin{array}{ll}        
   E^{-\beta}, & E>E_0 \\
   E^{-\beta - 1/\theta}, & E_1 < E < E_0 \\
   \const, & E < E_1 
   \end{array}\right.,
\end{equation} 
i.e. have two power-law parts the harder of which reflect the 
high-energy tail of the elementary spectrum and the softer accounts for the 
elementary spectrum evolution.  

\item{}
Let us now examine the case of a stepwise elementary spectrum 
(Fig.~\ref{Puchok}b) described by 
a Heaviside $\Theta$-function: $n(E) = \Theta(E_0 - E)$,
where $E_0$ evolves as in the previous example.
Then the integral spectrum should be
\begin{equation}
 N(E) = \left\{\begin{array}{ll}
  0, & E>E_0 \\
  \sim E^{-\frac{\sigma+1}{\theta}} , & E_1 < E < E_0 \\
  \const,  & E < E_1 
   \end{array}\right.,
\end{equation} 
i.e the Heaviside step is smoothed into a power function. The discontinuity
of the above $N(E)$ at the point $E_0$ is an artifact of the approximation:
in fact, there is not an exact power function, but very close to it if
$t_1 \gg t_0$ as supposed.

\item{} So we have shown that the integration of both  power and
stepwise spectra  leads to a power-law behaviour between $E_1$ and $E_0$.
The elementary spectrum with Plank (Wien)  high-energy tail 
 (Fig.~\ref{Puchok}c) lies between the power and stepwise cases,
it is not as steep as the stepwise one but steeper than the power one.
So it is natural to expect a similar result (power-law behaviour) 
for the integral spectrum.

\end{enumerate}

\subsection{The integration of the blackbody elementary spectrum}

Now we have come to the time integration of the blackbody Plank spectrum.
\begin{equation}
 n(E,t) = A(t){E^2 \over \exp[E/T(t)]-1} \; .
\end{equation}
Here we measure $E$ and $T$ in the same units, say, $T_0$, and
let us measure time in units of $t_0$, so instead of 
\begin{equation}
A(t)=A_0\left({t\over{t_0}}\right)^{\sigma},
\quad T(t)=T_0\left({t\over{t_0}}\right)^{-\theta}
\end{equation}
we have simply                                     
\begin{equation}
A(t)=A_0 t^\sigma, \qquad T(t)=t^{-\theta}.
\end{equation}

The observed integral spectrum  is
\begin{equation}
N(E)=\int_1^{t_1} dt
 A(t){E^2 \over \exp[E/T(t)]-1} =
 A_0\int_1^{t_1} dt { t^\sigma E^2 \over \exp(E t^\theta)-1} \; .
\end{equation}
Introducing $y=E t^\theta$, we rewrite this as
\begin{equation}
N(E)=
 A_0{E^{2-(\sigma+1)/\theta}\over \theta}
\int_E^{E t_1}   dy { y^{(\sigma+1)/\theta-1} \over \exp(y)-1} \; .
\end{equation}

From this general expression we derive the asymptotic cases.

\begin{enumerate}
\item The most interesting case is when $E<1$, that is $E<kT_0$ in
standard units, and $Et_1 \gg 1$. One should remember that $t_1$ is always
greater than unity, so the latter inequality is true when $E$ is not too
small. Then we find, replacing the lower integration limit by zero, and the 
upper one by infinity,
\begin{equation}
N(E) \simeq 
 A_0{E^{2-(\sigma+1)/\theta}\over \theta}\int_0^{\infty} 
  dy { y^{(\sigma+1)/\theta-1} \over \exp(y)-1} \; .
\end{equation}
The value of the integral is not interesting for us now.
Thus, we produce a power-law spectrum with the exponent $2-(\sigma+1)/\theta$.
Say, for $\sigma=2$ and $\theta=3/4$ we find the spectrum 
$N(E) \sim \nu^{-2}$.
For $\sigma=2$ and $\theta=1$ we find the spectrum $N(E)\sim E^{-1}$ 
(flat $F_\nu \sim \nu^0$), 
etc. See the numerical examples below.

\item When $E \ll 1$ and $E t_1 \ll 1$, we have $y \ll 1$, 
so we are in the Rayleigh-Jeans (RJ) regime, $\exp(y)-1 \simeq y$, and
\begin{equation}
 N(E) \simeq 
 A_0{E^{2-(\sigma+1)/\theta}\over \theta}
 \int_\nu^{E t_1}   dy  y^{(\sigma+1)/\theta-2}  \propto E \; .
 \label{NE_RJ}
\end{equation}
\item For high frequencies, $E>1$ and $E t_1 >> E >1$, 
the flux reduces to
\begin{equation}
N(E) \simeq 
 A_0{E^{2-(\sigma+1)/\theta}\over \theta}
\int_E^{E t_1}   dy  y^{(\sigma+1)/\theta-1}  e^{-y} \sim
 A_0{E^{2-(\sigma+1)/\theta}\over \theta} E ^{(\sigma+1)/\theta-1}
 e^{-E} \propto E^1 e^{-E} \; .
\end{equation}
So, here, in the Wien regime, for any $\sigma$ and $\theta$ we have in standard units
$N(E)  \sim E^1 \exp(-E/kT_0)$ [contrary to
$N_b(E) \sim E^2 \exp(-E/kT_0)$ for the blackbody of temperature $T_0$].

\end{enumerate}

\subsection{Numerical Examples}
\begin{figure}
\epsfysize=13cm
\centering{\epsfbox{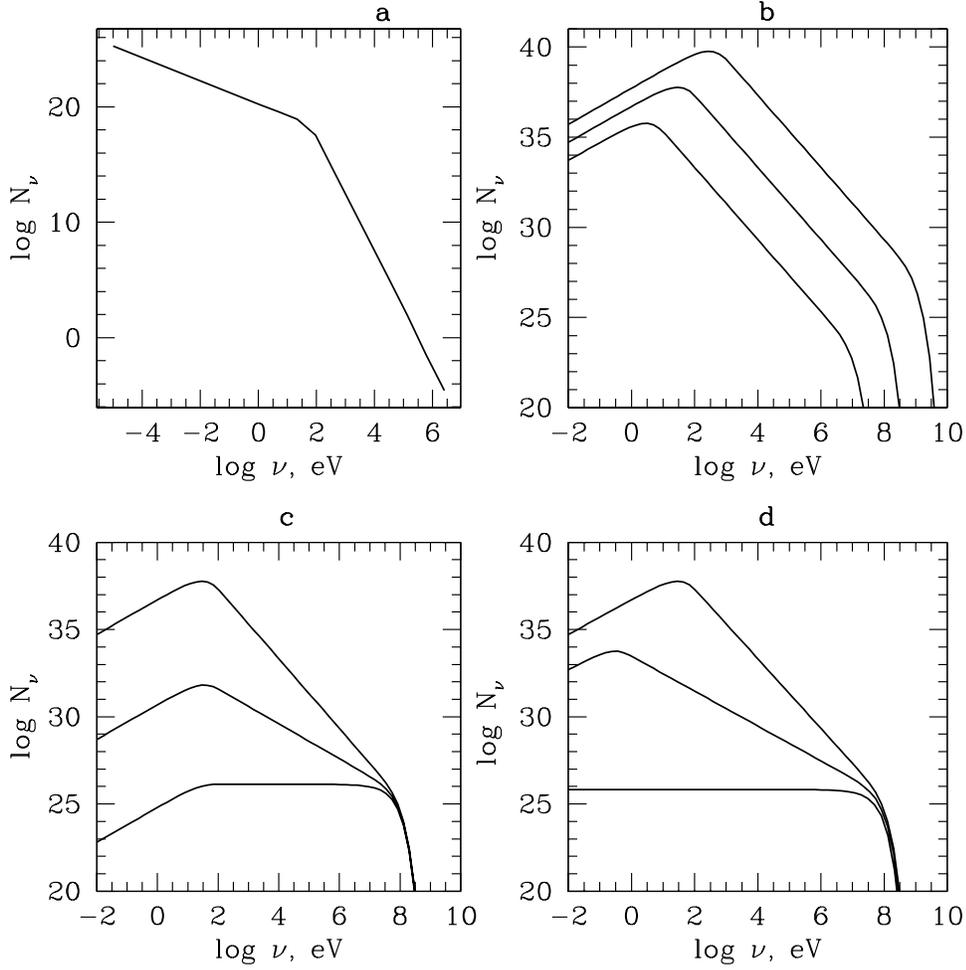}}
\caption{
The integrated spectra:
 a) step function with  
 $T_0=10^8$eV, $\sigma=2,\; \theta=0.75$; \hskip 2mm
 b) blackbody with $T_0={10^7,\;10^8,\;10^9}\mbox{eV},
          \quad\sigma=2,\;\theta=0.75$;  \hskip 2mm
 c) the same for  $T_0=10^8$eV, $\sigma={2,\;1.25,\;0.5},\;\theta=0.75$; \hskip 2mm
 d) the same for $T_0=10^8$eV, $\sigma=2$, $\theta={1.5,\;1,\;0.75}$
         }
\label{Mammoth}
\end{figure}

Fig.~\ref{Mammoth} presents the results of numerical integration 
of the $4$ cases of elementary spectra with various model parameters.
It also illustrates the correctness of the above analytical estimates.
One can compare these spectra with fig.~4 from  Chiang \& Dermer (1998)
where a similar time integration is done, but in a different physical
situation.

For a fixed pair of $\sigma$ and $\theta$ the spectrum consists of two power
laws and one exponential (Wien) high-energy part.
Therefore, it is in some sense similar to the Band (1993) 
function which also has two power law parts, so it can be expected to fit
the observations as well.

The moderately high energy part ($E_1<E<E_0$) has  the power law spectrum with the
the exponent depending both on the cooling ($\theta$) and expansion ($\sigma$) laws.
The dynamical range, i.e. the spectral width of this part is
$E_1/E_0 = (t_1/t_0)^{-\theta} \gg 1$.
Of course it depends on the integration time $\tau = t_1 - t_0 \approx t_1$,
and should be smaller for high temporal resolution. This can be a serious test
of the present model.

The highest energy part ($E>E_0$) represents the exponential breakdown, which
may be observed or not, depending on the value of $E_0$. Also, for such energies,
there may exists some other (optically thin?) radiation mechanisms which can
provide more intense emission  than the proposed blackbody one.

The low energy ($E<E_1$) part of the spectrum in our model should have only one
possible value of the slope. This is clear from the analytical
considerations: $\alpha=1$ (see eqn.~\ref{NE_RJ}).
This $\alpha$ is close to be consistent with many 
observations (Crider et al. 1997), 
but the observed variety of spectra
is much richer, than the simple RJ case, and there are claims that some GRB's
do show here the spectra predicted by synchrotron model (Cohen et al. 1997).
We can demonstrate, that with a small sophistication our blackbody model can
reproduce those spectra as well.

\begin{figure}
\epsfysize=8cm
\centering{\epsfbox{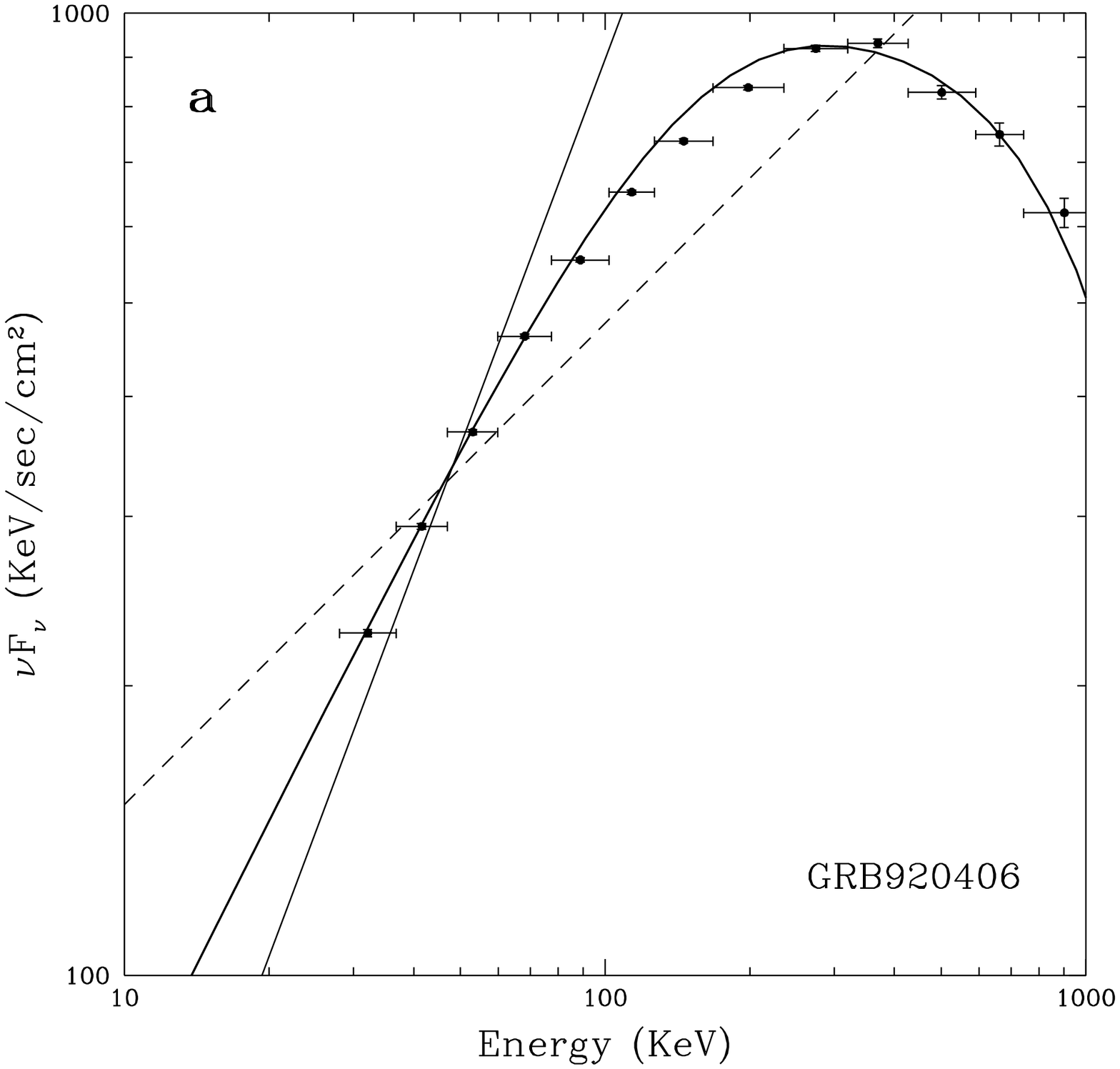}}
\epsfysize=8cm
\centering{\epsfbox{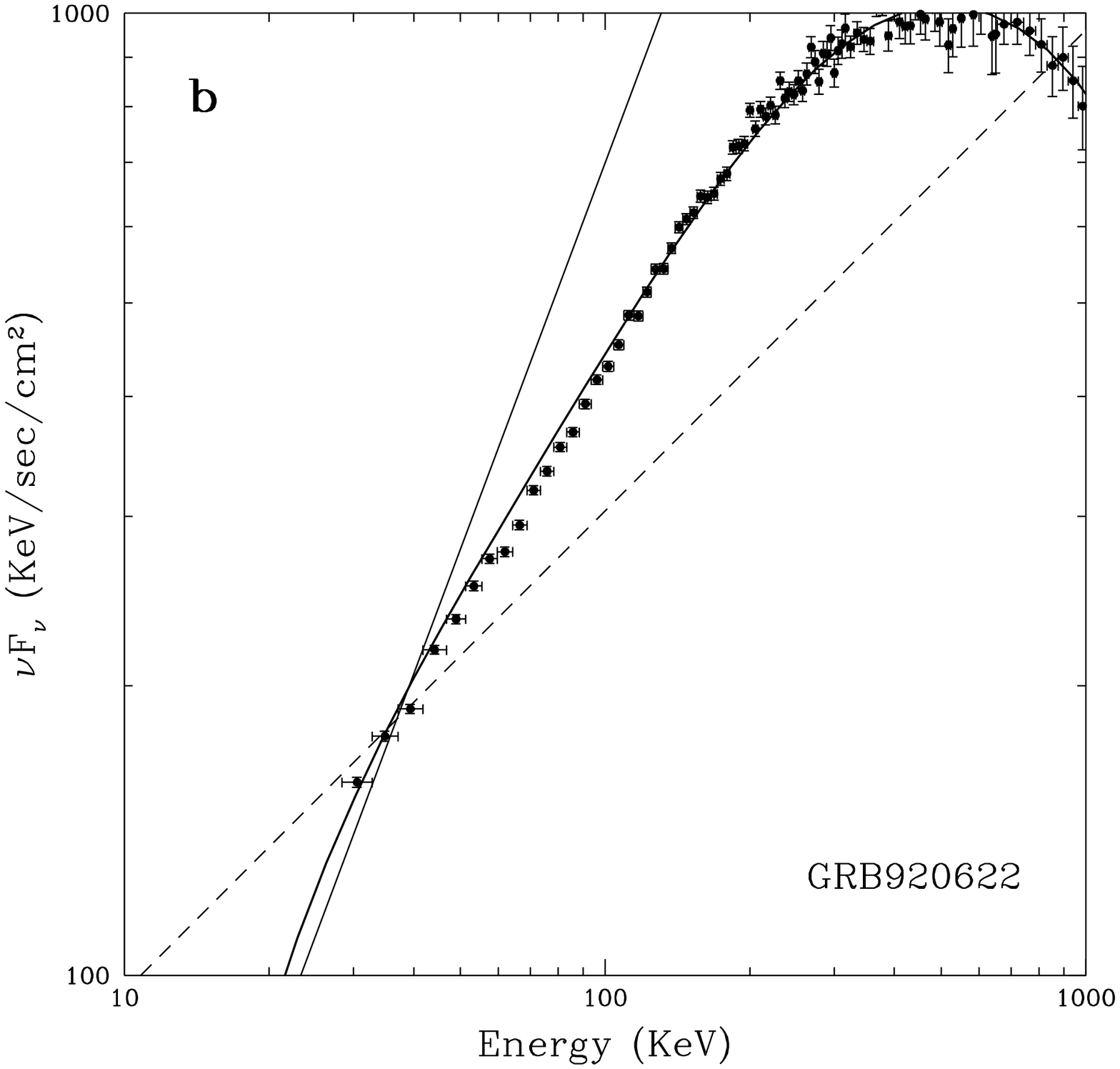}}
\caption{ a) GRB920406: $\theta=0.75,\;\theta_s=1.0,\;\lg T_0=5.45,\;t_1/t_0=
  1.2\cdot 10^3,\;f_{\rm hard}=0.0115$;
b) GRB920622: $\theta=0.75,\;\theta_s=0.955,\;\lg T_0=5.62,\;t_1/t_0
  =1.7\cdot 10^2,\;f_{\rm hard}=0.05$ }
\label{fit1}
\end{figure}

\begin{figure}
\epsfysize=8cm
\centering{\epsfbox{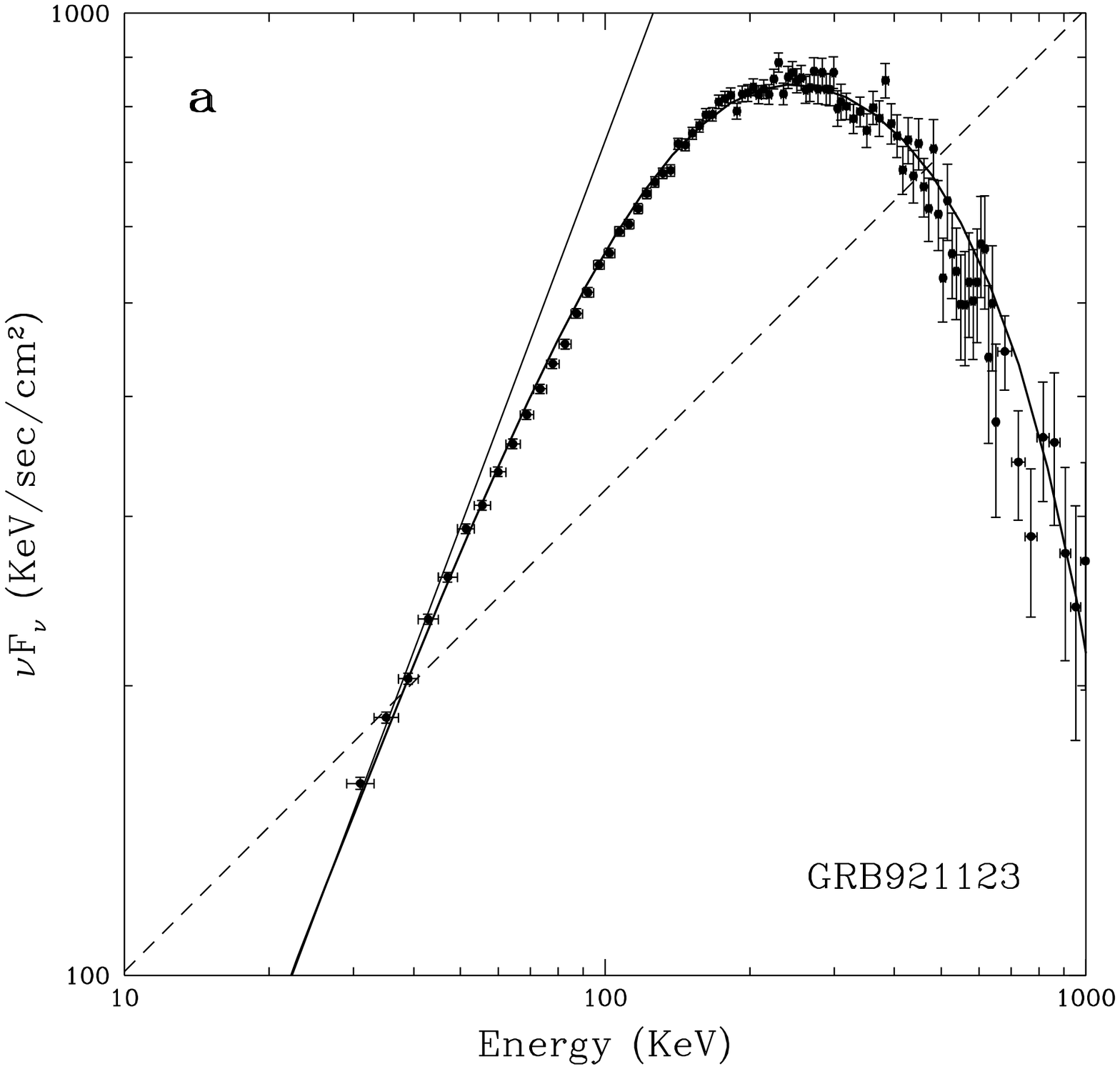}}
\epsfysize=8cm
\centering{\epsfbox{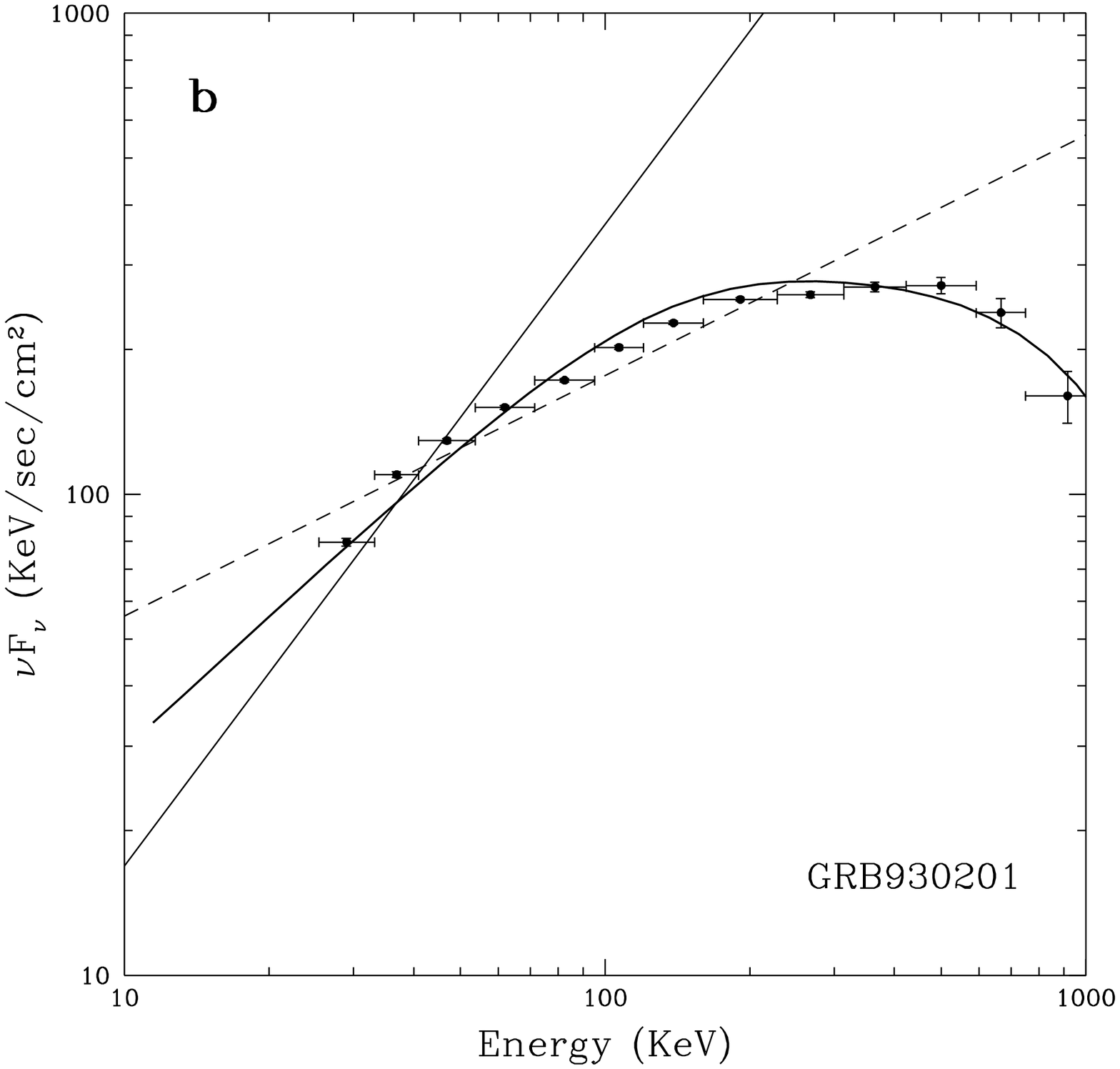}}
\caption{ a) GRB921123: $\theta=0.75,\;\theta_s=1.125,\;\lg T_0=5.28,\;t_1/t_0
=8.7\cdot 10^2,\;f_{\rm hard}=0.01$
          b) GRB930201: $\theta=0.75,\;\theta_s=0.975,\;\lg T_0=5.47,\;t_1/t_0
=1.3\cdot 10^3,\;f_{\rm hard}=0.014$}
\label{fit2}
\end{figure}

\begin{figure}
\epsfysize=8cm
\centering{\epsfbox{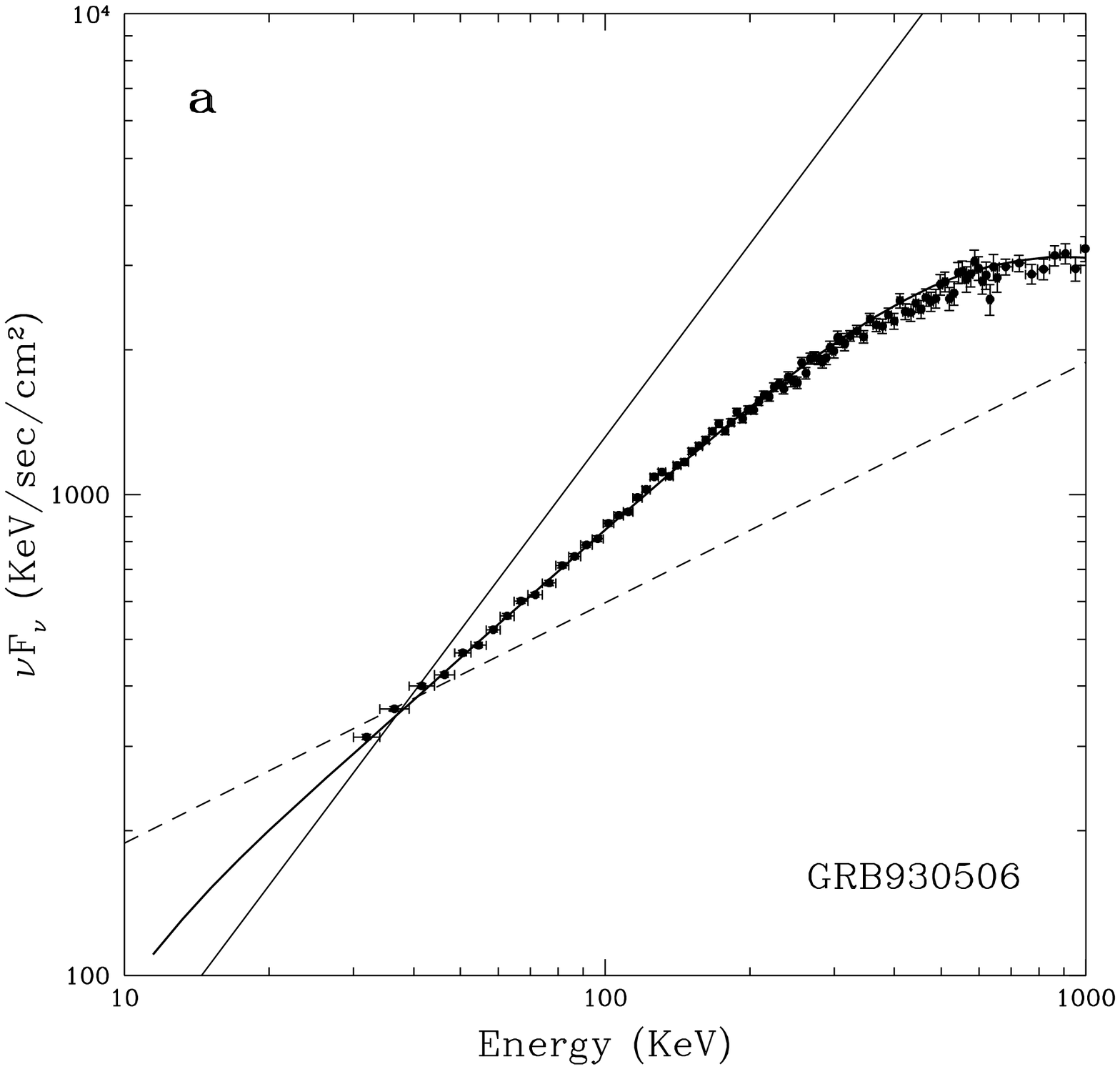}}
\epsfysize=8cm
\centering{\epsfbox{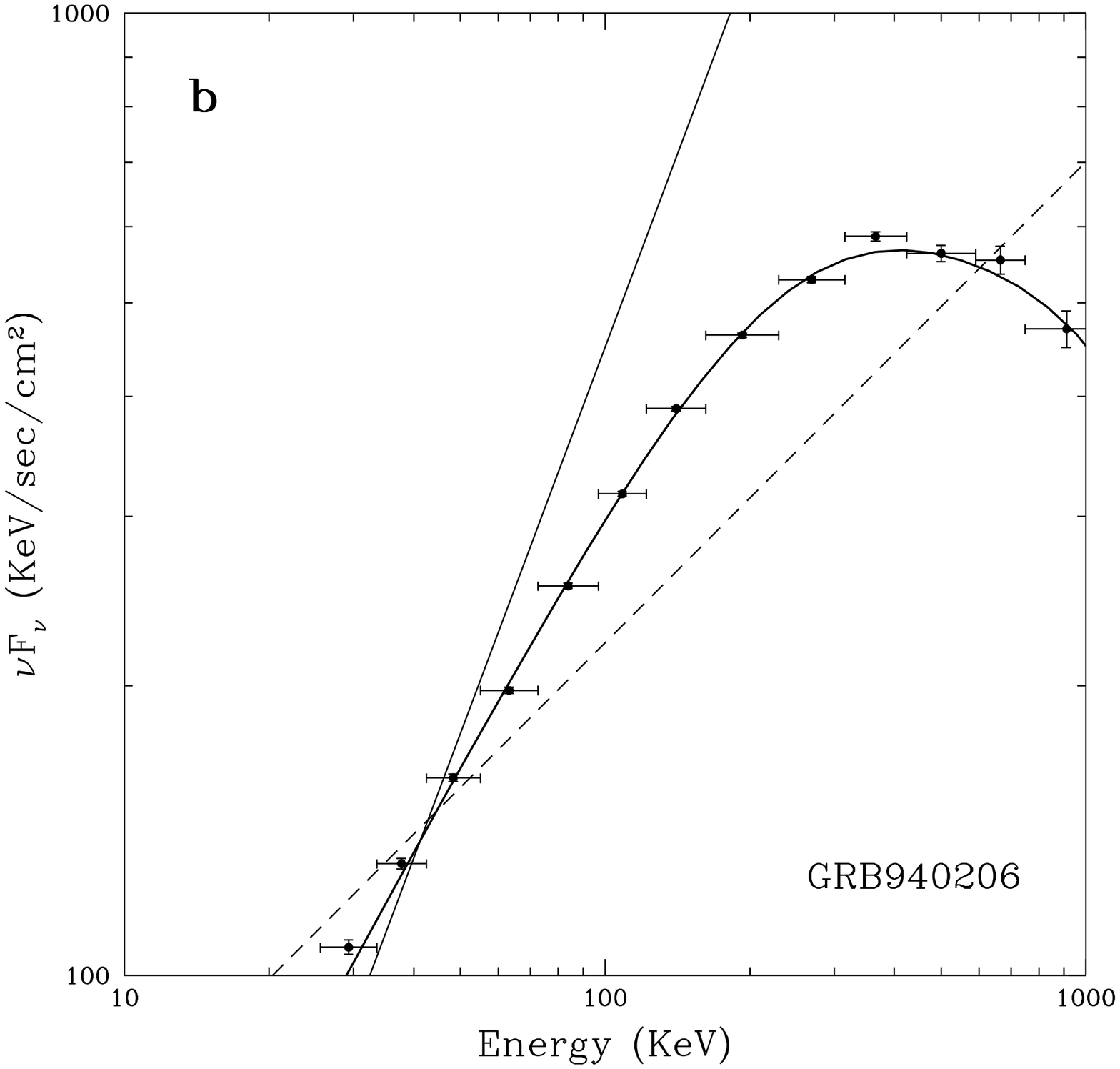}}
\caption{ a) GRB930506: $\theta=0.74,\;\theta_s=0.967,\;\lg T_0=6.0,\;t_1/t_0
=1.1\cdot 10^3,\;f_{\rm hard}=0.014$
          b) GRB940206: $\theta=0.74,\;\theta_s=0.975,\;\lg T_0=5.65,\;t_1/t_0
=1.1\cdot 10^3,\;f_{\rm hard}=0.014$}
\label{fit3}
\end{figure}

\begin{figure}
\epsfysize=8cm
\centering{\epsfbox{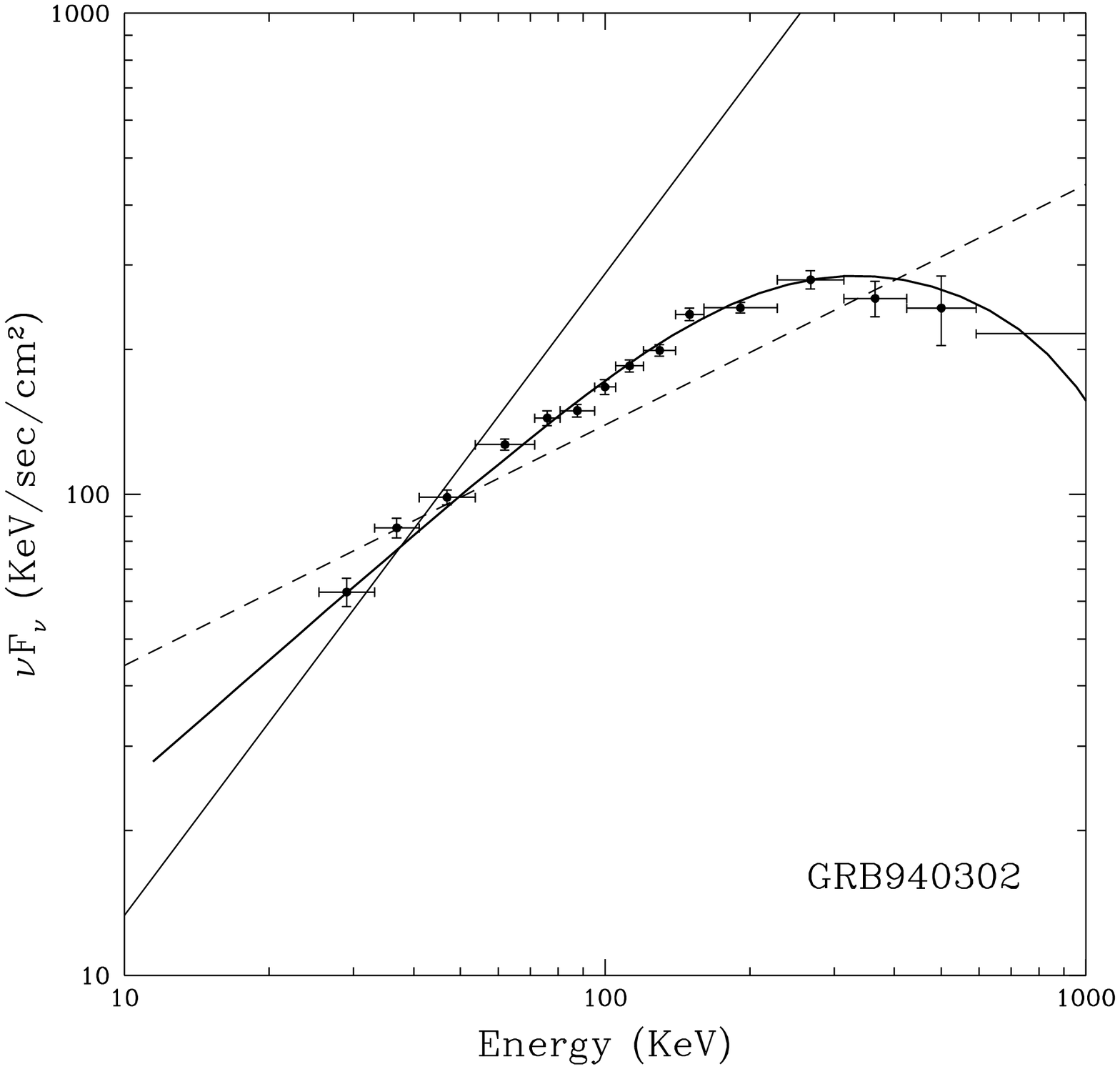}}
\caption{ GRB940302: $\theta=0.75,\;\theta_s=0.96,\;\lg T_0=5.45,\;t_1/t_0
=7.8\cdot 10^2,\;f_{\rm hard}=0.012$}
\label{fit4}
\end{figure}

Introducing new parameters is usually a means to improve a fit, but also makes
the latter physically less reliable. In what follows, we will keep one 
parameter constant, let us take $\sigma=2$, as the most natural choice. 
Instead, we can introduce a physically motivated additional 
parameter $f_{\rm hard}$ as the fraction of the time when the value of $\theta$
is constant, assuming that after some time, $f_{\rm hard}\tau$, the  temperature
power law changes. In the examples below, for the fixed $\sigma$ at constant 
value $2$, we  allow the value of $\theta$ to change a bit.
For illustration, we have taken GRB spectra given in (Cohen et al. 1997) 
in the form of postscript files and superimposed them onto our fits.
The spectra from (Cohen et al. 1997) 
are all integrated in time, it would be better
to have time-resolved spectra. But in any case it is not possible to
have time resolution better then 1 ms and for the illustration of our idea the
spectra used are quite good.
As shown by Figs. ~\ref{fit1},~\ref{fit2},~\ref{fit3},~\ref{fit4}, 
our black body model can provide good fits for 
the GRB spectra which were claimed to give evidence for synchrotron radiation.

\section{Discussion and Conclusions}

We have assumed that at each moment the spectrum of the 
gamma-ray burst emission is close to the black body one.
After the integration in time over the typical temporal resolution of the 
observations it produces a spectrum which can be similar to the observed
`non-thermal' GRB spectrum. In reality,
both the instantaneous spectrum and its true time evolution can deviate
significantly from our simplified assumption. So in reality one can have
a much richer variety of observed spectra.
In our work, we wish only to
point out a simple fact: that the observed non-thermal spectrum can be 
produced by an optically thick expanding body under fairly natural assumptions

We have in mind the following picture. The central engine of GRB
operates on a space scale like $10^6$ cm (the size of a neutron star
or a stellar mass black hole). It produces shells or
bullets of matter moving with the speed which is only one millionth slower
than the light speed $c$, i.e. we assume $\Gamma \sim 10^3$. The high value of
$\Gamma$
is needed in any case for cosmological GRBs in order to solve the compactness
problem (see e.g. Piran 1998
for all refs). But the standard picture invokes
the high  $\Gamma$ in order to make the fireball transparent only if it is clean
(without baryon load): they have the optical depth 
$\tau_{\gamma\gamma}$ going down  $\sim  \Gamma^{4+\beta}$, 
if the $\beta \sim 2$ is the index of the power 
spectrum at the hard tail $N(E) \propto E^{-\beta}$. 

So in standard
picture, and in our picture as well, if we see a pulse 
of GRB lasting $1$~ms, the size of the shell should have grown from $10^6$~cm
up to $10^6$~light milliseconds $=10^3$~light seconds $\sim 10^{14}$~cm,
since $R \sim  2\Gamma^2 c \times 1$~ms.
Now one can only  start speculating, where do the next pulses of GRB come
from. These can be internal or external shocks (Piran 1998), 
or light reflections (Shaviv \& Dar 1995, Drozdova \& Panchenko 1997), 
etc. However, there are arguments (Fenimore, Ramirez, \& Sumner 1997) that one
shell expanding forever is not able to produce GRB pulses which only show
a slight `hard to soft' evolution for hundreds of pulses. Already for the first
pulse the shell had to expand from $\sim 10^6$ cm to $\sim 10^{14}$ cm. 
Therefore, a model
where a central engine repeats shooting shells or bullets for the whole
duration of the GRB is preferred (see also Dar 1998). 

\begin{figure}
\begin{center}   
\epsfysize=8cm
\epsfbox{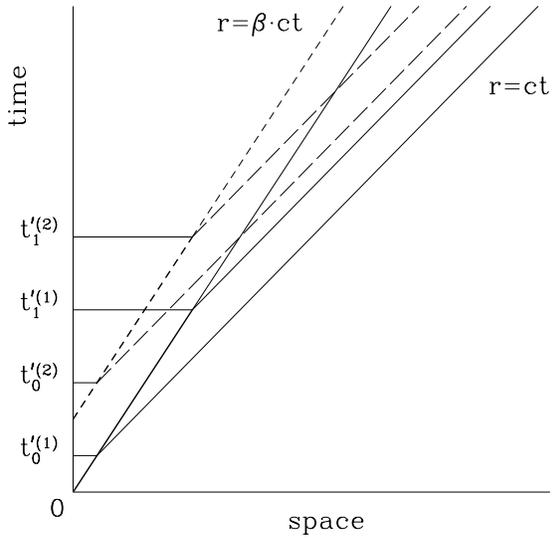}
\end{center}
\caption{
The space-time diagram for the emission of two shells.
Though the second shell is shot when the first one remains optically
thick, the observer sees the radiation of these two shells separately 
because of their ultra-relativistic velocities.
}
\label{lorents2}
\end{figure}

Thus, we have $R$ like $10^6$ cm and  $t_0 \sim 3\cdot 10^{-5}$ sec, and
$t_1$ (or $\tau$) like $10^3$ second, so $7$ orders of magnitude for the 
dynamical range of a power-law spectrum in our model is quite plausible.
This covers the range from keV to GeV. The evidence for hard, TeV, emission
associated with GRBs remains inconclusive (Padilla et al. 1998).
Only if the extremely hard TeV photons are detected, as suggested by
Totani (1998), then one should invoke a truly non-thermal emission
mechanism. We do not say that our shell has the thickness in the
end like $\sim 10^{14}$ cm, it must be geometrically thin, so more dense -- it
is {\em optically thick}. But its radius $R$ is of course $\sim 10^{14}$ cm, with
$dR\ll R$. It can be {\em loaded\/} with baryons to some extend, not violating
 the energy limits of course (Krolik \& Pier 1991). 
This is good if one has
something like stripping the surface layers of neutron stars (Blinnikov et
al. 1984; Eichler et al. 1989; Ruffert et al. 1997).
Reaching the
size $\sim 10^{14}$ cm our shell (or bullet) has expanded and cooled enough
to become transparent in the end. The shell traveled this distance $10^3$ seconds
according to our clocks, but one should not forget that it kept running almost
with speed of light, the light that it had produced. So the difference in time
of the beginning of the flash, that we see on Earth, and its end is only 1
millisecond.
While our shell is still very near the centre, the engine has shot already
the second shell (or bullet, Fig.~\ref{lorents2}), then the 3rd one, ...,
the 100th, etc. If the
total GRB duration observed on Earth was a few seconds, {\em all} the shots
of the central engine were done when our first shell was like (few seconds/$10^3$
seconds) smaller than in the end, so its radius was like
$\sim 10^{11}$ to $\sim 10^{12}$ cm. At this time it was very optically thick,
but one should not forget that it moves so fast,
that the light of the 2nd, 3rd, ..., 100th etc. shells can reach the first
shell only after the first one is far away, $\sim 10^{14}$ cm from the centre,
and absolutely transparent.

If instead of shells we have bullets, moving at some small angles to us
there is no problem of transparency. They can cool down and become small
solid bodies (this is perhaps not probable, since they must be heated 
up by ISM).

In reality, not only time, but also space integration takes place.
As shown by Rees (1966), (see also Drozdova \& Panchenko 1997, Sari 1998)
in the case of an expanding emitting shell an observer simultaneously detects
radiation produced in different moments of time (thus, with different
temperatures) on the ellipsoidal or egg-like surface.
The integration over this surface can 
give the same effect as the integration over time done in this paper,
but we do not perform this here because the result strongly depends on the
unknown geometry of the emitting surface. 

To conclude, we found that a variety of observed `nonthermal' GRB spectra
can be well reproduced by the time-integrated emission of a black-body
spectra. 
The most critical test of our model can be the discovery of the 
temporal resolution dependence of the power spectrum range
(here $E_1\dots E_0$). However, it can be smoothed by a space integration.
The main advantage of the proposed model is that  it 
allows the baryon load to be limited not by
the optical thickness, but by energy considerations only (one cannot
accelerate too much baryons because of their high rest mass).

{\bf Acknowledgements.}
Our work  is partly supported by RBRF grants 96-02-16352 and
96-02-19756, INTAS `Thermonuclear Supernovae', ISTC 97-370, and Russian
Federal programs `Astronomy' and `Science Schools'.
The work of IEP was made possible by the INTAS 96-0315 and  RFBR 98-02-16801
grants. Part of the work was done while SIB was visiting
Stockholm  Observatory under the grant of the Swedish Royal Academy of
Sciences, 
and he is grateful to Peter Lundqvist and Claes 
Fransson for their hospitality, to Felix Ryde for his data on GRBs, and to
Claes-Ingvar Bj\"ornsson for stimulating comments. The support of MPA,
Garching, and encouragement by Wolfgang Hillebrandt are gratefully
acknowledged.

\end{document}